\documentclass[a4paper,12pt]{article}
\usepackage{amsmath,amssymb}

\begin{document}
\begin{center}
\section*{The  cosmic  string  as a channel for the massive particle teleportation}

{S.V. Talalov}

\vspace{5 mm}

\small{Department of Applied Mathematics, State University of Toliatti, \\ 14 Belorusskaya str.,
 Toliatti, Samara region, 445020 Russia.\\
svt\_19@mail.ru}

\end{center}


\begin{abstract}
Here we prove the existence of a new type of the world-sheet string singularities - the cusps that are stable during the finite time. These singularities make  the emission of the captured  massive  quantum particle possible in the frames of the author's model suggested earlier. 
In aggregate, we have a new mechanism of quantum teleportation of such particles at large distances.

\end{abstract}

 {\bf keywords:}  {cosmic strings; cuspidal points.}
 
 {\bf PACS codes:}  03.65 Ge;  11.27 +d
 
 {\bf MSC codes:} 51P99;  53B50;    81V19;  85A99

\vspace{5mm}

     We investigate the  infinite $4D$  Nambu - Goto (NG) string  (see \cite{BarNes}, for example) that might be interpreted as a cosmic string in wire approximation\,\cite{Ander}. The cuspidal points arising during the string evolution\,\cite{Turok,Vil} lead to some interesting physical effects: for example, the gravitational wave bursts\,\cite{DamVil} and the radio bursts\,\cite{Vach} in the   Universe. One of such effects could be associated with the capture of        
      the massive quantum particle  and its subsequent  transfer   along the string.  The possibility of this   was demonstrated  by the author in the simple non-relativistic model recently\,\cite{Tal1}, where the  infinite $4D$ NG string $X_\mu(t,s)$ was considered in the gauge $X_0 \propto t$.  The ''spatial part''  of the 4-vector $X_\mu(t,s)$  -- the curve ${\bf x}(t,s) \in E_{3}$  was considered as a source of short-range potential   forces acting on the massive non-relativistic quantum particle, for  
every time $t$. The time-dependent  potential is defined by the matrix elements:

 \begin{equation}
\langle \bf p \mid \hat V_a(t) \mid \bf p' \rangle =
     {\epsilon_a}\, \chi_a(\bf p) \chi_a(\bf p')
\int\limits^\infty_{-\infty} e^{ -i( \bf p - \bf p' ) \bf x(t,\,s) } g(s)ds\,,
\label{separ_s}
\end{equation}
 where  vector $\bf p$ is  momemtum of a particle, the coupling  constant ${\epsilon_a} <0$ and the
  Heaviside function $\chi_a(\bf p) = \theta(1/a - |\,\bf p|)$. 
   The constant $a\sim 0$ restricts the domain of the considered forces to a small neighborhood of the curve ${\bf x}(s,t)$.
 The ''form-factor'' $g(s)$ is the arbitrary function from the Swarz space ${\mathcal S}$  that satisfies the conditions
 $g(s)\equiv 1$ $\forall\,s\in [-R,R\,]$ (for some  $R>>1$) and $0\le g(s)\le 1$.
 The relevance of this   ''separable'' approximation was justified in\,\cite{Tal1}. 
   The following effect was demonstrated: the particle, captured at any point ''A'' of the string,  can be 
 ''transferred'' to the other point ''B''  of the curve ${\bf x}(t,s)$ if the point ''B''  becomes the cuspidal point due to the string evolution.
 The standard situation when the cusps on the $4D$ string world-sheet form  isolated points only (see\,\cite{KliNik} for example).
 Thus, we have the collapse of a wave function of the captured particle caused by  the appearance of the cusp.   Because such point  disappears instantly,
 we have the effect of the mass transfer along the string only:  there are no any physical reasons for the emission of the captured particle  out  from a neighbourhood of the string.

In this article we demonstrate that the cusps  on an NG string    can exist  during the finite time.  To prove it we use the author's approach for the geometrical description of string
that has been developed in a number works (see\,\cite{Tal2,Tal3,Tal_Nova} fo details).
Let us describe the string dynamics (both regular and cuspy configurations)in Minkowski space-time $E_{1,3}$ in terms of suggested approach briefly.
The standard
procedure (see \cite{BarNes} for example) leads to the  dynamical equations
$\partial_{+}\partial_{-}X_\mu =0\,$   and constraints $(\partial_{\pm}X_\mu)^2 = 0$, where the derivatives
$\partial_\pm = \partial/\partial \xi_\pm$ and cone parameters $\xi_\pm =s\pm t\,$.
So,  the objects of our consideration will be time-like world-sheets with orthonormal parametrization.
 We consider the infinite strings, that's why we must impose
 asymptotic conditions on a curve $X_\mu = X_\mu(s)$ for any value of evolution parameter $t$.
 The demanded conditions  will be formulated  later; in the first place  we'll  point out to the local structure of the world-sheet.
Let us define  the pair of light-like  and scale-invariant vectors in space $E_{1,3}$:
 \begin{equation}
 \label{e_pm}
 {\bf e}_{\pm}(\xi_\pm) =  \pm \frac{1}{\varkappa} \,\partial_{\pm}{\bf X}(\xi_\pm)\,,
  \end{equation}
 where the value $\varkappa$ is an arbitrary (and scale transformed) positive constant.
In our opinion, the separation of scale-transformed mode is quite natural here because the scale invariance of the NG theory.
 It is clear that we can  construct  the pair of orthonormal bases
 ${\bf e}_{\nu\pm}(\xi_\pm)$ in space $E_{1,3}$  that are connected with the introduced vectors  ${\bf e}_{\pm}$
 by the equalities
 ${\bf  e}_{\pm}=\left({\bf e_{0\pm}} \mp {\bf e_{3\pm}} \right)/2$.
 Obviously, the definition of these  bases  has three - parameter arbitrariness in each
point $(t,s)$. We will keep  this fact in mind and eliminate
this ambiguity in the correspondent place.
The bases ${\bf e}_{\nu\pm}(\xi_\pm)$   allow us to define the vector-matrices  ${\hat{\bf E}_\pm}$:
 \begin{equation}
{\hat{\bf E}_\pm}= {\bf e_0}_\pm {\boldsymbol{1_2}} - \sum_{i=1}^3{\bf e_i}_{\pm}{\boldsymbol{\sigma_i}}\,.
     \label{matr_E}
\end{equation}
The basis ${\bf e}_-$ can be transformed into the basis ${\bf e}_+$  by the Lorentz transformation; that is why
we can define  the $SL(2,C)$ - valued field $K(t,s)$  by means of the formula:
\begin{equation}
{\hat{\bf E}_+} = K{\hat{\bf E}_-}K^{+}\,.
     \label{field_K}
\end{equation}
The field  $K(t,s)$ is important object for our approach.
In accordance with the definition of the vector-matrices ${\hat{\bf E}_\pm}$, these matrices satisfy the equalities $\partial_{\pm}
{\hat{\bf E}_\mp}=0$.  Consequenly, the matrix field $K(t,s)$  satisfies  to (special) WZWN - equation
\begin{equation}
\partial_+\left(K^{-1}\partial_-K\right)=0\,.
     \label{eq_K}
\end{equation}

Let us define the comlex-valued functions $\varphi(t,s)$ and
$\alpha_\pm(t,s)$ by means of Gauss decomposition for the
matrix $K(t,s)\in SL(2,C)$:
\begin{equation}
K = \left(\begin{matrix}
1&0\\
{-\alpha_+}&1
\end{matrix}\right)
\left(\begin{matrix}\exp(-{\varphi}/{2})&0\\
0&\exp({\varphi /2})\end{matrix}\right)
\left(\begin{matrix}1&\alpha_-\\
 0&1\end{matrix}\right)\,.
     \label{gauss}
\end{equation}
In general, these functions   are singular because the decomposition
(\ref{gauss}) is not defined for the points where the principal
minor $K_{11}=0$. Let us introduce the regular functions
$\rho_{\pm}=(\partial_{\pm}\alpha_{\mp})\,{\exp}(-\varphi)$. The
consequence of the equality (\ref{eq_K}) will be  the following PDE
- system:
\begin{subequations}
\label{TL_rho}
     \begin{eqnarray}
     \label{1_sys}
     \partial_+\partial_-\varphi &=& 2\rho_+\rho_-\exp\varphi,\\
     \label{2_sys}
     \partial_\pm\rho_\mp &=& 0,\\
     \label{3_sys}
     \partial_\pm\alpha_\mp &=& \rho_\pm\exp\varphi.
     \end{eqnarray}
\end{subequations}

From the geometrical viewpoint, the introduction of the function $\varphi$ and the functions $\rho_{\pm}$ is justified by the following formulae for the first  $({\bf I})$
and the pair of  second (${\bf II}_i$, $i=1,2$) fundamental forms of the world-sheet:
\begin{eqnarray}
{\bf I}& = &\partial_{+}X_\mu \partial_{-}X^\mu  d\xi_+d\xi_-  = -\frac{\varkappa^2}{2}\,{\rm e}^{-{\rm Re}\varphi}d\xi_+d\xi_-\,,\\
\label{form2}
{\bf{II}}_1+ {\rm i}{\bf{II}}_2  & = & \varkappa\left(|\rho_+(\xi_+)|e^{{\rm i}[\beta +\chi]}d\xi^2_+   -
|\rho_-(\xi_-)|e^{{\rm i}[\beta -\chi]}d\xi^2_-\right)\,,
\end{eqnarray}
In the formula (\ref{form2}) the function
$\beta\in [0, 2\pi)$ is an arbitrary parameter and the function  $\chi  =({\rm Im}\varphi + \arg\rho_+ + \arg\rho_-)/2$. These objects will not discussed here.  Of couse, the equations (\ref{1_sys}) and (\ref{2_sys}) could be deduced from the Gauss and Peterson-Kodazzi equations.  
The cuspidal points correspond to the singularities of the function  $\varphi$.
These singularities arise when the light-like vectors  ${\bf e}_{\pm} $  
coinside.  Therefore, the vector  $ \dot{\bf X} = \varkappa\,({\bf e}_{+}  +     {\bf e}_{-})$ be light-like vector when the 
condition  $\exp(-{\varphi}/{2})=0$ is fullfiled.

The reconstructing formulae for the space components of the  vectors ${\bf e}_{\pm} \equiv \pm (1/\varkappa\,)\, \partial_{\pm}{\bf X}(\xi_\pm)$  are:

\begin{equation}
 \Bigr(\,     -
{\rm Re}\left( \,t_{i1{\pm}}\overline{t}_{i2{\pm}}\right)\,, \quad   
  - {\rm Im}\left( \,t_{i1{\pm}}\overline{t}_{i2{\pm}}\right) \,,\quad  
\frac{1}{2} \,(- |t_{i1{\pm}}|^2 +|t_{i2{\pm}}|^2\,)\,\Bigl)\,, 
\label{e+}
 \end{equation}
 where the index $i$  is correlated with the sign $\pm$ in accordance witn the rule $ i=\frac{3\mp1}{2}$ and the  
 functions $t_{ij\pm}$ are the matrix elements  of certain  matrices $T_\pm(\xi) \in SL(2,C)$. These matrices solve the linear systems
\begin{equation}
     T^{\,\prime}_\pm(\xi)+
Q_\pm(\xi)T_\pm(\xi) = 0 \,,
     \label{spect1}
     \end{equation}
where
\begin{equation}
\label{Q_def}
Q_-(t,s) = K^{-1}\partial_-K\,,\quad Q_+ (t,s) =- (\partial_+K)K^{-1}\,.
\end{equation}
Let us note that transformation
$$ T_\pm(\xi) \longrightarrow {\widetilde T}_\pm(\xi)  =  T_\pm(\xi)B\,,\qquad B \in SL(2,C)\,,\qquad B_{ij} = const\,,$$
corresponds to the Lorentz transformation of space $E_{1,3}$.

Thus, we can  reconstruct the tangent vectors  ${\bf e}_{\pm}$ through the   solutions of 
the system (\ref{TL_rho}).
To do it we must calculate the coefficients $Q_{ij\pm}$ of the systems (\ref{spect1}) through
 the functions $\varphi$, $\rho_\pm$, $\alpha_\pm$ and  add a certain finite number of constants that
fix the solutions $t_{ij\pm}$. The singular functions $\varphi$ and $\alpha_\pm$ correspond to cuspy string configurations.

The system (\ref{TL_rho}) has a wide group ${\sf G}$ of invariance.
Indeed, let the functions $\varphi(\xi_+,\xi_-)$,  $\rho_\pm(\xi_\pm)$
and $\alpha_\pm(\xi_+,\xi_-)$
 be solutions for the system (\ref{TL_rho}). Then the transformation
\begin{equation}
     (\varphi, \rho_\pm,             \alpha_\pm)\longrightarrow
(\tilde\varphi, \tilde\rho_\pm, \tilde\alpha_\pm),
     \label{group_G}
     \end{equation}
     gives the new solution for the system (\ref{TL_rho}) if
     \begin{eqnarray}
  \tilde\varphi(\xi_+,\xi_-)&=&\varphi(A_+(\xi_+),A_-(\xi_-))+
     f_+(\xi_+)+f_-(\xi_-),\nonumber\\
  \tilde\rho_\pm(\xi_\pm)&=&
\rho(A_\pm(\xi_\pm))A_\pm^{\prime}(\xi_\pm)\exp{(-f_\pm(\xi_\pm))},
     \nonumber\\
\tilde\alpha_\pm(\xi_+,\xi_-)&=&
\alpha_\pm(A_+(\xi_+),A_-(\xi_-))\exp{(f_\pm(\xi_\pm))}
+g_\pm(\xi_\pm).\nonumber
     \end{eqnarray}
     for arbitrary complex-valued functions $f_\pm(\xi)$,  $g_\pm(\xi)$ and such real functions
    $A_\pm(\xi)$ where the conditions $A_-^{\prime}A_+^{\prime}\not= 0$ are fulfilled.
Let the subgroup $ {\sf G}_0\subset {\sf G}$  is defined by the conditions $A_\pm(\xi)\equiv \xi$.
Performing the factorization procedure for   the group $ {\sf G}_0$, we can to eliminate the uncertainty that arose early when the bases  ${\bf e}_{\nu\pm}(\xi_\pm)$ were defined.
Each coset of the group $ {\sf G}_0$ contains the   matrices $Q_\pm$ which have the form
\begin{equation}
\label{Q_red}
Q_\pm(\xi) =  -{\rho_\pm}(\xi)\boldsymbol{\sigma_\pm} + {\overline\rho}_\pm (\xi)\boldsymbol{\sigma_\mp}\,.
\end{equation}
Cosequently, the  corresponding matrices $T_\pm(\xi) \in SU(2).$ For our subsequent considerations we will take into account these representatives only.
We suppose the functions $\rho_\pm \in {\mathcal S}$; therefore,
the world-sheet asymptotically  is a planar surface  because (\ref{form2}).
After factorization procedure\,\footnote{which is non-covariant gauge fixing}  we have for the ''time'' component  of the 4-vector ${\bf X}(t,s)$:  $X_0(t,s) \equiv X_0^0 + \varkappa\,t$.

Finally, we have the description of the string dynamics in terms: 1) the  ''internal'' variables $\rho_\pm(s\pm t)$ which are invariant under rotations of the space $E_3$, space and time translations and scale transformations; 2) the scale transformed constant $\varkappa$; 3) the certain finite number of constants ($X^0_\mu$, \dots) that define the immersion of the variables
$\rho_\pm $ in space and time $E_3 \times R_1$.
In addition to a clearer description of the geometry, the suggested approach leads to additional possibilities for the hamiltonization of string dynamics. This question is beyond the topic of this article. Relevant studies, as well as
the proofs and the details, can be found in the works\,\cite{Tal2,Tal3,Tal_Nova}.

We emphasize that  the functions $\rho_\pm(\xi)$ are dynamical variables in our approach. They can be arbitrary functions from the  space  ${\mathcal S}$.
The restrictions  such as  $|\rho_\pm| > 0$ lead to additional constraints for dynamics, moreover they make  the definition of ''small variations'' 
$\delta\rho_\pm$  ambiguous in the Swarz space. There are no similar constraints here; thus  the identities 
\begin{equation}
\label{ident1}
\rho_\pm(\xi) \equiv 0\,, \qquad \xi\in[a_\pm,b_\pm]\,, \qquad -\infty \le a_\pm < b_\pm \le \infty\,, 
\end{equation}
 should  take place for the certain functions $\rho_\pm(\xi)$.  Note  that  there are no conformal transformations
$\xi_\pm \rightarrow \widetilde\xi_\pm = A_\pm(\xi_\pm)\,$,   ($A^{\prime}\not= 0$), in this case   
which could  trivialize the coefficients of the forms (\ref{form2})  globally.
 As a complement, we note also  that the possibility  to have non-isolated zeros\,\footnote{which are ignored usually} for the functions $\rho_\pm(\xi)$ 
   is valid in the pseudo-Euclidean target space only.   It is true  because the  metrics on the  (time - like) world-sheet is pseudo-Euclidean. In the Euclidean space we can choose the comlex coordinates $\xi$ and
$\overline\xi$ to parametrize the ''world-sheet'', instead of the real  cone  parameters $\xi_\pm$ for the  pseudo-Euclidean case.
 As a consequence the complex-analytical function $\rho_+(\xi)$ and the complex-(anti)analytical function $\rho_-(\overline\xi)$ can have 
isolated zeroes only. This fact has an obvious illustration:  the gaussian curvature  for any  soap film in the space $E_3$ is either zero everywhere or non-zero everywhere.

As a final step of the procedure which outlined above, let us write out the formula for the explicit reconstruction of the world-sheet
${\bf X}(t,s)$:
\begin{equation}
\label{X_final}
{\bf X}(t,s) = {\bf X}^0  +\varkappa\,\left(\int_0^{s+t} {\bf e}_{+}(\eta)d\eta \, + \int_0^{s-t} {\bf e}_{-}(\eta)d\eta\right)\,. 
\end{equation}
The space components of the light-like vectors ${\bf e}_{\pm}$ are defined by the formulae (\ref{e+}) and the time components are equal to $1/2$ for our gauge.
The singularities of the world-sheet  -- cusps -- correspond to the zeroes of the fundamental form ${\bf I}$: in terms of the functions $t_{ij\pm}$ we have
the following equality 
\begin{equation}
\label{dXdX}
\partial_{+}{\bf X} \partial_{-}{\bf X}  = -\frac{\varkappa^2}{2}\Big\vert \, t_{11+}(\xi_+)t_{22-}(\xi_-) -  t_{12+}(\xi_+)t_{21-}(\xi_-) \Big\vert^{\,2}\,.
\end{equation}
Thus the dynamics of a string cusps is defined by the  singular solutions of the system (\ref{TL_rho}).
So, the special case $\rho_\pm\equiv const$   of the equation (\ref{1_sys}) is the Liouville equation. The real  solutions (that correspond to the string in $3D$ space-time)
 with   singularities of the  Liouville equation  has been   investigated  firstly in the work \cite{JPP}.
The system  (\ref{TL_rho}) has been investigated  firstly in the work\,\cite{PogrTal} as a model of interacting scalar and spinor field in two-dimensional space - time. Corresponding singular solutions have been investigated in the work\,\cite{TrLv}.

Using the formula (\ref{X_final}),  we are going to prove that a string configuration  providing  to the conditions (\ref{ident1}),
can lead to the cusp that will be stable during the finite time. Indeed, the identities $T_\pm(\xi)\equiv T_{\pm}^0 = const$~
  $\forall\, \xi\in[a_\pm,b_\pm]$ are true for the case (\ref{ident1}) in general. The corresponding domain of the world-sheet
is a part of a time-like plane.
To simplify the consideration we suppose that $a_\pm =-b$, $b_\pm = b$ in the formulae (\ref{ident1}) for the  certain constant $b>0$. 
What happens if the equality 
\begin{equation}
\label{ident2}
T_{-}(\xi_-)\equiv   T_{-}^0 =-i\boldsymbol{\sigma_2}T_+^0 \equiv -i\boldsymbol{\sigma_2}T_{+}(\xi_+) \, \qquad |\xi_\pm| \le b\,,
\end{equation}
 is valid?  In this case the identities $ e_{\mu-}(\xi_-) \equiv e_{\mu+}(\xi_+) \equiv  e_{\mu} $ are true for the certain constant light-like vector  $e_{\mu} $.
  The r.h.s. of the equality (\ref{dXdX}) is equal to zero identically for  all $s,t$ such that   $  |s\pm t| \le    b$.
 Thus, if the time $''t''$ satisfies the unequality
 $$ -b +|s| \le t \le  b -|s|\,,  $$
 for the certain values of the parameter $s$, the world-sheet degenerates into the light-like line segment
 \begin{equation}
\label{wsh-deg}
  X_\mu (s,t)  = X_\mu^0   +  \varkappa\, t\,e_{\mu}\,,
  \end{equation}
 where the vector $X_\mu^0$ is a constant vector.
  Thus, the cusp on the string $X_\mu(\cdot,s)$ that arises at the moment $t_0 = -b$  in the point $X_\mu(-b,0)$, will be stable during the period $\Delta t = t_1 - t_0$, 
 before the moment $t_1 = b$.
 Note that the s-parametrization the curve $X_\mu(\cdot,s)$ for the moments $t_0 < t <t_1$ is degenerated: we have the identity
 $\partial X_\mu/\partial s\equiv 0$ for all $s$ that satisfies inequality $-b+|t| < s < b - |t|$.  This fact means that the parameter $s$ may not be the length of the arc of the curve
$X_\mu(\cdot,s)$ in a singular case in general.

We can ease the condition  (\ref{ident1}): let the the identity 
\begin{equation}
\label{ident3}
{\rm Im}\,\rho_\pm(\xi) \equiv 0\,, \qquad \xi\in[a_\pm,b_\pm]\,, \qquad -\infty \le a_\pm < b_\pm \le \infty\,, 
\end{equation}
is fulfilled only.
In accordance with the formulae (\ref{e+}),  (\ref{spect1}) and   (\ref{Q_red}) the considered string becomes planar on the domain $\xi_\pm\in[a_\pm,b_\pm]$. 
As it is well-known the planar strings can have  stable cusps.

What does the stability of the cusp mean from the viewpoint of the model\,\cite{Tal1} that demonstrates the capture of a massive particle and it transfer  into the cuspidal point? 
In our opinion the cusp that will be stable during the finite time provides not only collapse of the wave function but the subsequent emission of this particle out from string too.
Indeed, the cusp moves with the velocity of light $c$, any massive particle moves with the velocity $v<c$. Therefore,  if the cusp exists during the finite time, the cuspidal point and the point of the location of the considered particle must diverge with probability  $P$ where $0< P < 1$.
Of course, the corresponding quantity theory for emission effect must be relativistic.  We hope to make  a more definite conclusions about the value 
$P = P(a, \Delta t, \dots)$ in the subsequent works.

This research did not receive any specific grant from funding agencies in the public, commercial, or not-for-profit sectors.

{\bf Acknowledgements.}
I would like to thank A.K. Pogrebkov who drew my attention   to the certain  ''pathological'' 
 singularities\,\cite{Pogr}     in the field model\,\cite{PogrTal} that motivated me for 
the  research presented here .

\end{document}